\documentclass[%
 reprint,
superscriptaddress,
 amsmath,amssymb,
aps,
floatfix,
]{revtex4-2}

\usepackage{graphicx}
\usepackage{dcolumn}%
\usepackage{bm}%
\usepackage{color}
\definecolor{navyblue}{rgb}{0, 0, 0.5}
\usepackage[colorlinks=true, breaklinks=false, linkcolor=navyblue, urlcolor=navyblue, citecolor=navyblue]{hyperref}
\newcommand{\fref}[2]{\hyperref[#1]{\ref*{#1}#2}} %
\usepackage[all]{hypcap}%
\usepackage[per-mode=symbol,separate-uncertainty]{siunitx}
\usepackage{braket}

\begin{document}

\title{%
Quantum-circuit refrigeration of a superconducting microwave resonator well below a single quantum}

\author{Arto Viitanen}
\email[Corresponding author.\\]{arto.viitanen@aalto.fi}
\affiliation{QCD Labs, QTF Centre of Excellence, Department of Applied Physics, Aalto University, P.O. Box 13500, FI-00076 Aalto, Finland}
\author{Timm Mörstedt}
\affiliation{QCD Labs, QTF Centre of Excellence, Department of Applied Physics, Aalto University, P.O. Box 13500, FI-00076 Aalto, Finland}
\author{Wallace S. Teixeira}
\affiliation{QCD Labs, QTF Centre of Excellence, Department of Applied Physics, Aalto University, P.O. Box 13500, FI-00076 Aalto, Finland}
\author{Maaria Tiiri}
\affiliation{QCD Labs, QTF Centre of Excellence, Department of Applied Physics, Aalto University, P.O. Box 13500, FI-00076 Aalto, Finland}
\author{Jukka Räbinä}
\affiliation{IQM Quantum Computers, FI-02150 Espoo, Finland}
\author{Matti Silveri}
\affiliation{Nano and Molecular Systems Research Unit, University of Oulu, P.O. Box 3000, FI-90014 Oulu, Finland}
\author{Mikko Möttönen}
\email[Corresponding author.\\]{mikko.mottonen@aalto.fi}
\affiliation{QCD Labs, QTF Centre of Excellence, Department of Applied Physics, Aalto University, P.O. Box 13500, FI-00076 Aalto, Finland}
\affiliation{QTF Centre of Excellence, VTT Technical Research Centre of Finland Ltd., P.O. Box 1000, 02044 VTT, Finland}

%\date{\today}
\date{August 1, 2023}

\begin{abstract} %
We experimentally demonstrate a recently proposed single-junction quantum-circuit refrigerator (QCR) as an in-situ-tunable low-temperature environment for a superconducting 4.7-GHz resonator. With the help of a transmon qubit, we measure the populations of the different resonator Fock states, thus providing reliable access to the temperature of the engineered electromagnetic environment and its effect on the resonator. We demonstrate coherent and thermal resonator states and that the on-demand dissipation provided by the QCR can drive these to a small fraction of a photon on average, even if starting above 1~K. We observe that the QCR can be operated either with a dc bias voltage or a gigahertz rf drive, or a combination of these. The bandwidth of the rf drive is not limited by the circuit itself and consequently, we show that 2.9-GHz continuous and 10-ns-pulsed drives lead to identical desired refrigeration of the resonator. These observations answer to the shortcomings of previous works where the Fock states were not resolvable and the QCR exhibited slow charging dynamics. Thus this work introduces a versatile tool to study open quantum systems, quantum thermodynamics, and to quickly reset superconducting qubits. 
\end{abstract}

\maketitle

Superconducting quantum circuits~\cite{Blais21} have proven to be a promising platform for devising advanced quantum-information systems~\cite{Kjaergaard20,Arute19,Krantz19,Kim23}. At the very heart, the devices rely on fundamental understanding of the physics of open quantum systems~\cite{Breuer16}, which can potentially be unlocked by reservoir engineering~\cite{Harrington22}. A promising implementation of tunable dissipation has been recently realized as a quantum-circuit refrigerator (QCR)~\cite{Tan16,Silveri19} which was developed from the study of photon-assisted tunneling in circuit quantum electrodynamics~\cite{Devoret90}. Owing to the versatility of the device, the QCR can operate as a tunable source of dissipation for a spectrum of quantum-electric devices~\cite{Silveri17,Hsu20,Hsu21,Yoshioka21}, finding applications in both industry such, as in qubit reset~\cite{Sevriuk22,Yoshioka23,Magnard18}, as well as in fundamental physics, such as in observing curious Lamb shifts~\cite{Silveri19,Aiello22,Viitanen21}, engineering exceptional points~\cite{Partanen19,Teixeira23}, quantum thermodynamics~\cite{Pekola15}, and strong environment-induced non-linearity in the high-impedance regime~\cite{Esteve18}.

The recent generations of QCRs are implemented as double normal-metal--insulator--superconductor (NIS) junctions capacitively connected to the refrigerated quantum device. When operated as a QCR, the voltage-controlled NIS tunnel junctions introduce photon-assisted quasiparticle tunneling, absorbing photons from the coupled circuit which is thus refrigerated. Consequently, the QCR provides an environment with orders of magnitude of tunability in the resulting dissipation rate~\cite{Silveri17,Silveri19}. The use of two junctions stems from the need of a return path for the dc current. Although feasible in principle, this design leads to a charge island, which experiences highly non-linear charging dynamics owing to the high subgap resistance of the NIS junctions. In recent experiments~\cite{Sevriuk22}, this non-linearity lead to a highly non-exponential decay of a QCR-damped transmon qubit. Furthermore, the asymmetry in the junctions cannot be conveniently measured, and hence poses challenge in modeling the device.

To resolve the issue of the slow charging effect of the island, very recently a single-junction QCR was theoretically proposed~\cite{Vadimov22}. Here, a single NIS junction is shunted to ground by a quarter-wave microwave resonator which provides the return path to ground of the dc current. Such a device has no charge island at dc and is expected to exhibit clean exponential behavior for the energy decay of the refrigerated quantum-electric devices. However, no experimental demonstration of the single-junction QCR has been reported. In addition, in all previous experiments of refrigerating a linear resonator with a QCR, only average power or photon number has been extracted. Thus it remains open how close to a thermal state the refrigerated system has relaxed.

\begin{figure*}
    \includegraphics[]{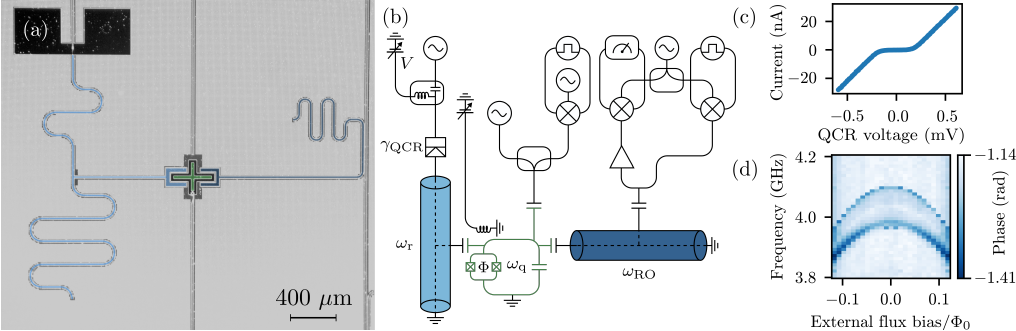} %
    \caption{Experimental device and setup. (a) False-color micrograph of a representative chip. The quarter-wave resonator (light blue) with an attached QCR at the top end is probed with a transmon qubit (green) setup consisting of a readout resonator (dark blue), a transmission line (right), a qubit drive line (top), and a flux line (bottom). (b) Simplified circuit diagram of the chip and the experimental setup with colors matching to those in (a). Each Josephson junction of the qubit is denoted by a crossed square, whereas the QCR junction has a half of a cross (superconductor side) and a blank rectangle (normal-metal side). (c) Current through the QCR as a function of its bias voltage $V$ without excitation applied elsewhere in the circuit. (d) Phase of the readout resonator signal as a function of the external flux bias $\Phi$ of the qubit and the excitation frequency of the qubit. Owing to a relatively high qubit-drive power, the two-photon transition from the ground state to the second excited state is visible below the primary ground-state-to-excited-state line.}
    \label{fig:setup}
\end{figure*}

\begin{figure}
    \includegraphics[]{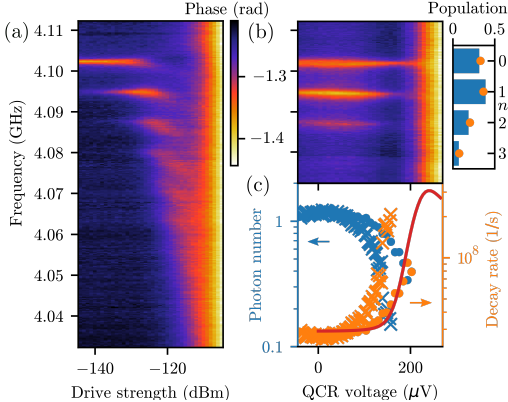} %
    \caption{Control of dissipation of a coherently driven resonator. (a) Phase of the readout signal as a function of the qubit excitation frequency and the power of a coherent drive tone at the resonator frequency. (b) Phase of the readout signal as a function of the qubit excitation frequency and the QCR bias voltage $V$ for the resonator driven at $-130$-dBm power. The right inset shows the measured populations $P_n$ (bars) of the Fock states $\ket{n}$ at $V=68~\mu$V and a fit to a Poisson distribution (dots) yielding $\bar n = \sum_n nP_n = 1.21$. (c) Mean resonator population $\bar n$ (blue dots, left axis) and the corresponding QCR-induced decay rate (orange dots, right axis) as functions of the QCR bias voltage extracted from the data of panel (b). The line shows our theoretical decay rate model $\gamma_\mathrm{QCR} + \gamma_\mathrm{V=0}$, for which some of the parameters are shown in Table~\ref{tab:parameters} and the superconductor gap parameter $\Delta = 220~\mu$eV, normal-metal temperature $T_\mathrm{N} = 150$~mK, junction capacitance $C_\mathrm{NIS} = 0.54$~fF, and resonator characteristic impedance $Z_\mathrm{r} = 63.7~\Omega$. The crosses show results of an identical experiment except with an additional continuous $9.25$-GHz drive applied at the junction.} %
    \label{fig:coherent}
\end{figure}

\begin{figure}
    \includegraphics[]{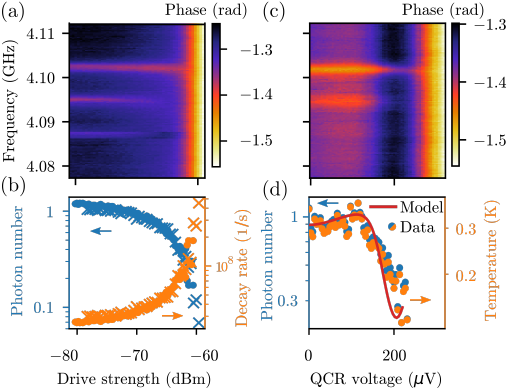} %
    \caption{Continuous and pulsed rf operated QCR and cooling of a thermal state. (a) Phase of the readout signal as a function of the power of a $10$-ns-pulsed rf excitation at the QCR and the qubit excitation frequency. The pulses have $50\%$ duty cycle and $2.9$-GHz carrier frequency. The resonator is coherently driven with $-130$-dBm power. (b) Mean photon number (blue markers, left axis) and corresponding QCR-induced decay rate (orange markers, right axis) for pulsed (crosses) and continuous (dots) $2.9$-GHz excitation as functions of the QCR rf tone strength. The QCR is not dc biased in panels (a) and (b). (c) Phase of the readout signal as a function of the QCR dc bias voltage and the qubit excitation frequency in the case of a hot attenuator. The attenuator is intentionally heated by a detuned rf signal with power tuned to have roughly a single thermal photon on average in the resonator. (d) Mean photon number (blue dots, left axis) and the corresponding effective temperature of the resonator environment (orange dots, right axis) as functions of the QCR bias voltage. The line shows our decay rate model with values as in Table~\ref{tab:parameters} and in Fig.~\ref{fig:coherent} except for $T_\mathrm{N} = 150$~mK, $\gamma_\mathrm{dr}/(2\pi) = 2$~MHz, and $\bar n_\mathrm{dr} = 1$.} %
    \label{fig:fig3}
\end{figure}

\begin{figure}
    \includegraphics[]{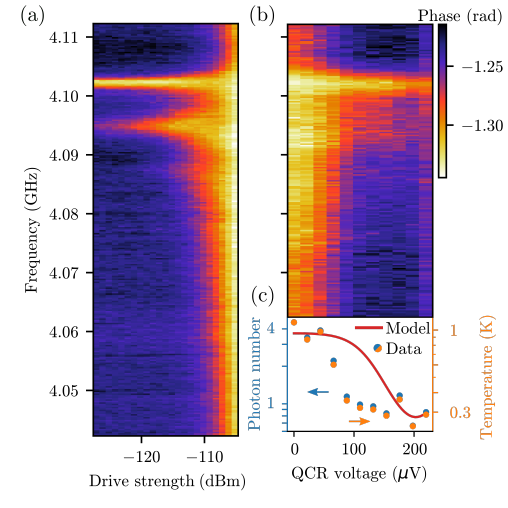} %
    \caption{Cooling of an intentionally heated resonator. (a) Phase of the readout signal as a function of the artificial thermal-noise power and the qubit excitation frequency. (b) Phase of the readout signal as a function of the QCR bias voltage and the qubit excitation frequency. The resonator is driven with artificial noise of approximately $-105$~dBm. (c) Mean resonator population and the corresponding temperature as function of the QCR bias voltage. The line shows our model with values given in Table~\ref{tab:parameters} and in Fig.~\ref{fig:coherent} except for $T_\mathrm{N} = 280$~mK, $\gamma_\mathrm{dr}/(2\pi) = 2$~MHz, and $\bar n_\mathrm{dr} = 4$.} %
    \label{fig:thermal}
\end{figure}

In this Letter, we implement a single-junction QCR~\cite{Vadimov22} galvanically connected to a quarter-wave superconducting resonator (Fig.~\ref{fig:setup}). The Fock states of the resonator are probed with a transmon qubit~\cite{Koch07} in a number-splitting experiment. Using the circuit we demonstrate in-situ voltage control of the dissipation on a coherent state of the resonator (Fig.~\ref{fig:coherent}). Importantly, we demonstrate control of the dissipation with both continuous and $10$-ns-pulsed rf signals of 2.9-GHz carrier frequency verifying the very broad operational bandwidth of the single-junction QCR. In another experiment, we intentionally heat up an attenuator and hence the resonator to a thermal state of a single photon on average. Subsequently, we refrigerate this thermal state to a small fraction of a single quantum (Fig.~\ref{fig:fig3}). In addition, we manage to heat the resonator by artificial high-temperature noise to the many-photon regime, corresponding to a roughly 1-K thermal state, and while applying this artificial heat, we cool the resonator with the QCR below a single quantum (Fig.~\ref{fig:thermal}). These advancements solve the previous major issue of slow charging of the QCR and introduce convenient tools to characterize the refrigerated states at the quantum level.

Our sample shown in Fig.~\fref{fig:setup}{(a)} consists of a quarter-wave superconducting coplanar-waveguide resonator, at the voltage antinode of which we galvanically connect the single-junction QCR. The resonator is capacitively coupled to a typical transmon qubit and its readout resonator is designed for the strong-dispersive regime.

The dc characteristics of the single-junction QCR are extracted from the current-voltage (IV) measurement using the setup illustrated in Fig.~\fref{fig:setup}{(b)}. The IV curve shown in Fig.~\fref{fig:setup}{(c)} allows us to extract the Dynes parameter $\gamma_\mathrm{D} \approx 9.25\times10^{-3}$ and tunneling resistance $R_\mathrm{T} \approx 14.7$~k$\Omega$ from the differential resistances within and far above the gap region~\cite{Karimi22}. %

Next, we find the resonance frequency of the readout resonator in a usual single-tone experiment, and then use the two-tone technique~\cite{Kjaergaard20} to measure the two lowest qubit transitions from its ground state as shown in Fig.~\fref{fig:setup}{(d)}. These data appears typical for a transmon qubit, which encourages us to proceed with more involved experiments.  %

In Fig.~\fref{fig:coherent}{(a)}, we illustrate a typical two-tone spectroscopy of the qubit transition frequency for various power levels of an additional continuous drive applied through the qubit drive line at the resonator frequency $\omega_\mathrm{r}/(2\pi) = 4.6704$~GHz. %
The qubit is operated at its flux sweet spot with a transition frequency of $\omega_\mathrm{q}/(2\pi) = 4.1024$~GHz at low-power resonator drive. At increasing power of the resonator drive, we observe several equidistant resonance peaks for the qubit with spacing $2\chi/(2\pi) = -7.45$~MHz. This peak spacing is in agreement with the ac-Stark shift of $2\chi\hat{a}^\dagger\hat a$ of the qubit, where $\chi$ is the resonator-qubit dispersive shift and $\hat a^{(\dagger)}$ is the resonator annihilation (creation) operator. 

Importantly, our device operates in the strong-dispersive regime, i.e., the dispersive shift is greater than the linewidths, $2\chi > \gamma,\kappa$, where $\gamma$ and $\kappa$ are the resonator and the qubit energy decay rates, respectively. Consequently, the individual spectral lines induced by the populations of each resonator Fock state can be resolved in Fig.~\fref{fig:coherent}{(a)}~\cite{Gambetta06,Schuster07,Blais21}. At high resonator populations, the spectral lines broaden until they are no longer resolvable and the device escapes the dispersive regime where the average resonator photon number $\bar n$ is much smaller than $n_\mathrm{crit} := \left[(\omega_\mathrm{q} - \omega_\mathrm{r}) / (2g)\right]^2 \approx 12$~\cite{Blais21}. Here, $g/(2\pi) = 80.7$~MHz is the qubit-resonator coupling strength obtained from the measured ac-Stark shift using the transmon anharmonicity of $-273$~MHz. However, at high populations the QCR becomes active owing to multi-photon tunneling processes and induces dissipation and Lamb shift~\cite{Viitanen21}.

\begin{table}[b]
    \caption{Summary of the experimentally measured device parameters.}
    \label{tab:parameters}
    \begin{ruledtabular}
        \begin{tabular}{lr}
            \textrm{Parameter, symbol} & \multicolumn{1}{c}{\textrm{Value}} \\
            \colrule
            Resonator frequency, $\omega_\mathrm{r}/(2\pi)$ & $4.6704$~\si{\giga\hertz}  \\
            Qubit frequency, $\omega_\mathrm{q}/(2\pi)$ & $4.1024$~\si{\giga\hertz} \\
            Readout resonator frequency, $\omega_\mathrm{RO}/(2\pi)$ & $7.4386$~\si{\giga\hertz} \\
            Qubit-resonator coupling strength, $g/(2\pi)$ & $80.7$~\si{\mega\hertz} \\
            Anharmonicity, $\alpha/(2\pi)$ & $-273$~\si{\mega\hertz}\\
            \vspace{.1in} %
            Dispersive shift, $\chi/(2\pi)$ & $3.725$~\si{\mega\hertz} \\
            Dynes parameter, $\gamma_\mathrm{D}$ & $9.25\times10^{-3}$ \\
            Tunneling resistance, $R_\mathrm{T}$ & $14.7$~\si{\kilo\ohm} \\
        \end{tabular}
    \end{ruledtabular}
\end{table}

The relative height of the $n$:th spectral line in Fig.~\fref{fig:coherent}{(a)} equals the probability $P_n$ of the resonator to lie in the corresponding Fock state $\ket{n}$. This allows us to distinguish between coherent and thermal states as well as to measure the mean photon number $\bar n$ in the resonator. In a coherent state, the resonator assumes Poisson statistics $P_n = e^{-\bar n}\bar n^n / n!$, whereas in a thermal state, we have a Gibbs distribution $P_n = \bar n^n / (\bar n + 1)^{n + 1}$~\cite{Schuster07}. These distributions are fitted to the measured spectra, which yields the mean resonator photon number $\bar n$ that we find to agree well with $\bar n = \sum_n nP_n$.

In order to accurately measure the effect of the QCR on the coherent state, we continuously drive the resonator through the qubit drive line while sweeping the QCR voltage (Fig.~\fref{fig:coherent}{(b)}). Once active, the QCR induces additional dissipation to the resonator resulting in a diminished mean photon number observed as disappearing spectral lines of the high-energy Fock states. For a more quantitative description of the coherent-state refrigeration, Fig.~\fref{fig:coherent}{(c)} shows the mean photon number extracted from the spectral line heights as a function of the QCR bias voltage. We observe the initial mean photon number of slightly above unity to fall to a fraction of a photon owing to the QCR.

At QCR voltages beyond the gap voltage $\Delta/e$, the spectral lines are no longer resolvable possibly due to QCR induced Lamb shifts~\cite{Silveri19}, excitations~\cite{Masuda18}, or refrigeration~\cite{Tan16} of various modes of the system. As a result of these phenomena, we claim that the fixed-frequency readout tone falls out of tune at high QCR voltages, causing a qubit-drive frequency independent background.%

Knowing the resonator drive power $P_\mathrm{r}$ and the resonator occupation $\bar n$, the energy decay rate of the resonator $\gamma$ is obtained from the power balance, $P_\mathrm{r} = \gamma\hbar\omega_\mathrm{r}\bar n$. However, only a part of the incident power is inserted into the resonator, depending on the resonator-drive line coupling strength $\gamma_\mathrm{dr}$. Nonetheless, the decay rate is accurately described by $\gamma=\gamma_\mathrm{QCR} + \gamma_\mathrm{V_\mathrm{QCR}=0}$, where the QCR-induced dissipation $\gamma_\mathrm{QCR}$ on a directly coupled resonator owing to single-photon tunneling processes is given by 
\begin{equation}
    \gamma_\mathrm{QCR} = \pi\frac{Z_\mathrm{r}}{R_\mathrm{T}}\sum_{\ell,\tau=\pm1}\ell\overrightarrow{F}(\tau eV + \ell\hbar\omega_\mathrm{r}),
\end{equation}
where $\overrightarrow{F}$ is the normalized forward quasiparticle tunneling rate, $V$ is the QCR bias voltage, and $Z_\textrm{r}$ is the characteristic impedance of the resonator mode. See Appendix for more details of the model based on Ref.~\cite{Silveri17}. We show in Fig.~\fref{fig:coherent}{(c)} that this model captures well the exponential rise of the resonator decay rate as a function of the QCR bias voltage.

In Ref.~\cite{Viitanen21}, the QCR was driven for the first time using rf excitation, namely, by exciting the second mode of the resonator where the QCR was embedded. Curiously, we add in Fig.~\fref{fig:coherent}{(c)} to the dc bias line an off-resonant $9.25$-GHz microwave tone through a bias tee%
, yet it shifts the increase of the QCR dissipation to lower dc voltages. The size of the observed effective voltage shift $\Delta V$ approximately corresponds to a single additional $9.25$-GHz photon absorbed in the tunneling process, $e\Delta V \approx\hbar\times 9$~GHz.

The above result of increasing the QCR refrigeration using an off-resonant microwave tone that can be turned on and off in the broadband transmission line extremely quickly, motivates us to apply pulsed microwaves to the QCR to demonstrate quick turn-on and turn-off of the dissipation. In Figs.~\fref{fig:fig3}{(a),(b)}, we show on a coherently driven resonator that the QCR operates essentially identically with a continuous $2.9$-GHz excitation as it does for a $10$-ns pulsed excitation with a $50\%$ duty cycle and equal average power%
, indicating the possibility of operating the island-free single-junction QCR with rf pulses as short as $10$~ns.

Let us next focus on the effect of the QCR on thermal states. Here, the QCR can be considered as decreasing the effective temperature $T$ of the electromagnetic environment, which is equivalent to the mean photon number through the Bose occupation $\bar n_\mathrm{thermal} = 1/\left\{\exp{[\hbar\omega_\mathrm{r}/(k_\mathrm{B}T)]} + 1\right\}$. In Fig.~\fref{fig:fig3}{(c),(d)} the QCR is shown to refrigerate the resonator from above $300$~mK to close to $100$~mK. The initial thermal state is chosen for strong spectral lines implying clear signal and prepared by heating an attenuator using a detuned heater signal. The attenuator emits black-body radiation to the resonator, increasing the resonator population. %
In our experiments the resonator thermal distribution can be accurately measured down to $60$~mK.%

The temperature of the thermal resonator state follows the temperatures of the coupled baths, with the mean-photon number given by $\bar n = (n_\mathrm{QCR}\gamma_\mathrm{QCR} + n_\mathrm{dr}\gamma_\mathrm{dr}) / (\gamma_\mathrm{QCR} + \gamma_\mathrm{dr})$, where $n_\mathrm{QCR}$ is the population of the QCR environment obtained from its effective temperature
\begin{equation}
    T_\mathrm{QCR} = \frac{\hbar\omega_\mathrm{r}}{k_\mathrm{B}}\left[\ln\left(\frac{\overrightarrow{F}(eV + \hbar\omega_\mathrm{r})}{\overrightarrow{F}(eV - \hbar\omega_\mathrm{r})}\right)\right]^{-1},
\end{equation}
$n_\mathrm{dr}$ is that of the drive line, and $\gamma_\mathrm{dr}$ is the resonator-drive line coupling strength. Note that the QCR-induced bath can be effectively turned on or off at will since $\gamma_\mathrm{QCR, off} \ll \gamma_\mathrm{dr} \ll \gamma_\mathrm{QCR, on}$, a fact which we use to fit $n_\mathrm{dr}$ and $T_\mathrm{N}$ from the measured resonator photon numbers when the QCR is turned off and on, respectively. We use $\gamma_\mathrm{dr}$ as a free parameter. The model follows the data well but has noticeably steeper voltage response to the QCR-induced dissipation, which we expect to follow from the atypically strong smearing of the IV curve of the QCR junction.%

Finally, we investigate our device at elevated temperatures for the thermal resonator state. To overcome the issue of heating the cryogenic system owing to a very hot attenuator, we obtain such states by applying artificial thermal noise band-limited close to the resonator frequency. In Fig.~\fref{fig:thermal}, we observe that we can heat the resonator state close to $1$~K and that the QCR can cool such state down to as low as $250$~mK. The data does not seem to follow as closely the theoretical model as in the above results, which we attribute to unexpected drift in the QCR bias voltage.

In conclusion, we have implemented and experimentally demonstrated the operation of an island-free single-junction QCR with all dc, continuous rf, and pulsed rf voltages. The single-junction design simplifies the fabrication and operation of QCRs, and provides very broad bandwidth of the QCR operation voltage. Attached to a resonator and probed with a qubit utilizing a number-splitting experiment, the QCR is shown to refrigerate the resonator to well below a single quantum for both coherent and thermal states. In fact, this Letter demonstrates in-situ reduction of the effective environmental temperature by almost an order of magnitude. For scientific and industrial applications, our circuit design supports the reset of superconducting qubits~\cite{Sevriuk22} using advanced techniques, such as $\ket{f,0} \leftrightarrow \ket{g,1}$ transition~\cite{Magnard18} combined with pulsed QCR operation to empty the resonator~\cite{Yoshioka21,Yoshioka23}. To this end, we aim to add thermalization blocks to the QCR to achieve low electron temperatures~\cite{Feshchenko15} that was not necessary for this work. Perhaps even more interestingly, this work represents a major step forward implementing versatile engineered environments for control of open quantum systems and quantum thermodynamic systems. For example, we aim to implement in the future noise refrigeration of a resonator, where the active cooling of the resonator is achieved by powering the QCR by thermal noise photons~\cite{Viitanen21}. Such a device converts spurious thermal noise into useful refrigeration, which may lower the energy consumption of quantum-technological devices such as quantum processing units.

\begin{acknowledgments}
    This work was funded by the Academy of Finland Centre of Excellence program (project Nos. 352925, and 336810) and grant Nos.~316619 and 349594 (THEPOW). We also acknowledge funding from the European Research Council under Advanced Grant No.~101053801 (ConceptQ) and the provision of facilities and technical support by Aalto University at OtaNano—Micronova Nanofabrication Centre. We thank Vasilii Vadimov, Jukka Pekola, Bayan Karimi, Miika Rasola, Ilari M\"akinen, Satrya Christoforus, Arman Alizadeh, Gianluigi Catelani, Joachim Ankerhold, J\"urgen Stockburger, and Tapio Ala-Nissila for scientific discussions and Heidi Kivij\"arvi for technical assistance.%
\end{acknowledgments}

\appendix

\section{Theoretical model}

The QCR-resonator system is here modeled independently of the dispersively coupled transmon, which is used to only probe the quantum state of the resonator. Such a QCR-resonator case is analyzed in detail in Ref.~\cite{Silveri17} for a double-junction device. Following the derivations, we obtain the single-junction forward and backward transition rates between states $\ket{qm}$ and $\ket{q+1m'}$ at NIS junction bias voltage $V$ as
\begin{equation}
    \Gamma^{\pm}_{qmm'}(V) = M_{mm'}^2\frac{R_\mathrm{K}}{R_\mathrm{T}}\overrightarrow{F}(\pm eV + \ell\hbar\omega_\mathrm{r} - E_q^\pm),
\end{equation}
where $q$ is the normal-metal island charge state, the resonator Fock state changes from $m$ to $m'$, $M_{mm'}$ is the related transition matrix element, $R_\mathrm{K}$ is the von Klitzing constant, $\ell = m - m'$, $E_q^\pm = E_\mathrm{N}(1\pm2q)$ is the change of the normal-metal island charging energy $E_\mathrm{N}$, and $\overrightarrow{F}$ is the normalized rate of forward quasiparticle tunneling. According to Ref.~\cite{Silveri17}, we have 
\begin{equation}
    \overrightarrow{F}(E) = \frac{1}{h}\int \mathrm{d}\varepsilon\, n_\mathrm{S}(\varepsilon)[1 - f_\mathrm{S}(\varepsilon)]f_\mathrm{N}(\varepsilon - E),
\end{equation}
where $f_\mathrm{N}$ and $f_\mathrm{S}$ are the Fermi functions of the normal-metal and superconducting leads, respectively, and $n_\mathrm{S}$ is the normalized superconductor quasiparticle density of states~\cite{Dynes78}
\begin{equation}
    n_\mathrm{S}(\varepsilon) = \left|\operatorname{Re}\left\{\frac{\varepsilon + i\gamma_\mathrm{D}\Delta}{\sqrt{(\varepsilon + i\gamma_\mathrm{D}\Delta)^2 - \Delta^2}}\right\}\right|.
\end{equation}
Replicating the treatment of Ref.~\cite{Silveri17} for these equations, we obtain the resonator transition rates
\begin{equation}
    \Gamma_{mm'} \approx M_{mm'}^2\frac{R_\mathrm{K}}{R_\mathrm{T}}\sum_{\tau=\pm1}\overrightarrow{F}(\tau eV+\ell\hbar\omega_\mathrm{r}-E_\mathrm{N}),
\end{equation}
which for single-photon processes yields the coupling strength and effective temperature
\begin{gather}
    \gamma_\mathrm{QCR} = \pi\frac{C_\mathrm{c}^2}{C_\mathrm{N}^2}\frac{Z_\mathrm{r}}{R_\mathrm{T}}\sum_{\ell,\tau=\pm1}\ell\overrightarrow{F}(\tau eV + \ell\hbar\omega_\mathrm{r} - E_\mathrm{N}), \\
    T_\mathrm{QCR} = \frac{\hbar\omega_\mathrm{r}}{k_\mathrm{B}}\left[\ln\left(\frac{\overrightarrow{F}(eV + \hbar\omega_\mathrm{r} - E_\mathrm{N})}{\overrightarrow{F}(eV - \hbar\omega_\mathrm{r} - E_\mathrm{N})}\right)\right]^{-1},
\end{gather}
where $C_\mathrm{c}$ is the QCR-resonator coupling capacitance, and $C_\mathrm{N}$ is the normal-metal island capacitance to ground. For the lowest mode of the quarter-wave coplanar-waveguide resonator, the resonator impedance of the lumped-element model is obtained as $Z_\mathrm{r} = \frac{4}{\pi}Z_0$, where $Z_0$ is the characteristic impedance of the coplanar waveguide. In the main text, the QCR is directly attached to the resonator, a case that is here modeled by the limit of a large coupling capacitance, which also effectively removes the island charging energy. Thus we set $C_\textrm{c}/C_\textrm{N}=1$ and $E_\textrm{N}=0$ in the above equations. An identical model can be obtained by starting from the single-junction system Hamiltonian as was theoretically established in Ref.~\cite{Vadimov22}.

\section{Sample fabrication}

The samples are fabricated on six-inch prime-grade intrinsic-silicon wafers. The circuit is defined with optical lithography and etching a 200-nm sputtered Nb film. The nanostructures of the NIS and SQUID junctions are each defined by their own electron beam lithography (EBL) step, followed by a lift-off process in acetone. In EBL, we employ a bilayer resist mask consisting of poly(methyl methacrylate) and methyl methacrylate to create an undercut and enable two-angle shadow evaporation. Before evaporation, surface oxides are removed from the contact area by argon milling. For each junction, the first evaporated layer is Al which is in-situ oxidized before the evaporation of the top layer.

To improve the interface between aluminum oxide and Cu in the NIS junction, a 3-nm Al film is deposited on top of the aluminium oxide of the tunnel barrier before the final 60-nm layer of Cu is applied to form the normal-conducting electrode. Due to the inverse proximity effect, the 3-nm Al film remains a normal metal. 

The NIS junction is written into a $120 \times 100~\si{\nano\meter\squared}$ area, although the junction realized in fabrication is typically slightly larger. For estimating the junction capacitance we use a typical value of capacitance per area of $45~\si{\femto\farad\per\micro\meter\squared}$~\cite{Masuda18}.

\end{document}